\documentclass[a4paper,12pt]{article}
\pdfoutput=1 

\usepackage{amsmath,amssymb}
\usepackage{slashed}

\usepackage[pdftex]{hyperref}

\usepackage{cite}
\usepackage{color}
\usepackage{graphicx}

\makeatletter

\def\del{\partial}

\makeatletter
\g@addto@macro\bfseries{\boldmath}
\makeatother

\def\half{{\frac{1}{2}}}

\def\unit{{1\kern-.65ex {\rm l}}}
\def\1{{1\kern-.65ex {\rm l}}}

\def\be{\begin{equation}}
\def\ee{\end{equation}}

\newcount\hour \newcount\minute
\hour=\time \divide \hour by 60
\minute=\time
\def\now{%
\ifnum \hour<13
  \ifnum \hour=0 \advance \hour by 12 \number\hour:\else \number\hour:\fi%
     \ifnum \minute<10 0\fi%
     \number\minute%
\ A.M.%
\else \advance \hour by -12 \number\hour:%
  \ifnum \minute<10 0\fi%
  \number\minute%
  \ P.M.%
\fi%
}

\makeatother

\begin{document}

\baselineskip=18pt  
\numberwithin{equation}{section}  



\renewcommand*{\thefootnote}{\fnsymbol{footnote}}


\thispagestyle{empty}

\vspace*{2.cm} 
\begin{center}
 {\LARGE Revisiting constraints on uplifts to de Sitter vacua}\\
 \vspace*{1.7cm}
 {\large Nana Cabo Bizet$^{1,2,3,}$\footnote{nana@fisica.ugto.mx} and Shinji Hirano$^{1,2,}$\footnote{shinji.hirano@wits.ac.za}}\\
 \vspace*{1.0cm} 
{\it  $^1$ School of Physics and Mandelstam Institute for Theoretical Physics \& NITheP,\\
 $^2$ DST-NRF Centre of Excellence in Mathematical and Statistical Sciences (CoE-MaSS),\\
  University of the Witwatersrand,\\
  WITS 2050, Johannesburg, South Africa\\[1ex]
$^3$ Departamento de F\'isica, Universidad de Guanajuato, \\
 Loma del Bosque 103, CP 37150, Le\'on, Guanajuato, M\'exico}

\end{center}
\vspace*{1.5cm}

\noindent
We revisit the issue of uplifting the potential to de Sitter (dS) vacua in type IIB flux compactifications of Kachru, Kallosh, Linde and Trivedi (KKLT).
We shed light on some tension between two constraints on dS vacua in type IIB string theory. 
One is the well-known and much-discussed constraint which leads to the no-go theorem that can in principle be evaded. The other follows from 4-dimensional Einstein's equations, which has, however, been much less discussed in connection with the former constraint.
In addition to the challenges previously posed, it is suggested that the uplifting scenarios, in particular, obstruct the evasion of the no-go theorem more strongly than one might have assumed.

\newpage
\setcounter{page}{1} 





\section{Introduction}
\label{sec:Introduction}
\renewcommand*{\thefootnote}{\arabic{footnote}}\setcounter{footnote}{0}

The landscape of vacua in string theory has had a great impact on the way we perceive string theory as a potential  description of nature. It challenges the idea that dynamics spontaneously selects the universe we live in, neutralising the relevance of naturalness questions, and even predicts the existence of multiverse \cite{Linde:2015edk}. A logical consequence, the prospect of the anthropic principle \cite{Susskind:2003kw}, has generated heated debates for and against it within and outside the string theory community.
The existence of the landscape, `discretuum' of vacua, was first conceived and suggested by Bousso and Polchinski in \cite{Bousso:2000xa}, and it has come to fruition in the work of KKLT \cite{Kachru:2003aw} who, for the first time, put foward the construction of (Anti-)dS vacua with all moduli fixed, building on the preceded work by Giddings, Kachru and Polchinski (GKP) \cite{Giddings:2001yu}.\footnote{The type IIA counterpart of the KKLT construction was pioneered in \cite{DeWolfe:2005uu}. See also, for example, \cite{Grana:2005jc} for a comprehensive review.}

The KKLT construction of dS vacua starts from GKP's Minkowski vacua with all but K\"ahler moduli fixed and proceeds in two steps: (1) Adding D7- or Euclidean D3-branes wrapping a four-cycle in the CY 3-fold which induce nonperturbative (NP) corrections to the potential and in effect stabilise the K\"ahler moduli at AdS vacua and (2) adding anti D3-branes to break supersymmetry (SUSY) and uplift the minima of the potential to meta-stable dS vacua with all moduli fixed.
By taking these steps as opposed to all-at-once approaches, this uplifting scenario is advantageous in the sense that it clarifies the physical origin of the SUSY breaking while ensuring the meta-stability of dS vacua and thus makes the existence of the vast landscape plausible \cite{Douglas:2003um}.
It is a virtue of the illuminating two-step construction that a few simple appealing variants have been subsequently proposed:  
(1) Burgess, Kallosh and Quevedo (BKQ) replaced anti D3-branes in the second step by magnetised D7-branes which yield the D-term SUSY breaking and purportedly uplift the potential to meta-stable dS vacua \cite{Burgess:2003ic}.
(2) Balasubramanian and Berglund (BB) put forward an uplifting mechanism without SUSY breaking D-branes instead by the leading $\alpha'$ correction to the K\"ahler potential on top of the NP effects  \cite{Balasubramanian:2004uy} which was later refined by Westphal \cite{Westphal:2006tn}.
(3) Brustein and de Alwis advocated the racetrack model in which multiple NP potentials are added by turning on discrete Wilson lines to split wrapped D7-branes with gaugino condensation into multiple clusters,  
while manifest SUSY breaking D-branes are absent and $\alpha'$ corrections to the K\"ahler potential are negligible \cite{Brustein:2004xn}.
(4) BB, Conlon and Quevedo provided a systematic way to construct meta-stable dS vacua compactified on exponentially large volume Calabi-Yau manifolds. Both the racetrack potential and $\alpha'$ corrections to the K\"ahler potential are included in conjunction with the KKLT or BKQ uplift \cite{Balasubramanian:2005zx}.

In this paper we focus on the uplifting scenarios with manifest SUSY breaking D-branes, namely, KKLT's anti D3- and BKQ's magnetised D7-branes. 
Although these uplifting scenarios seem to be rather compelling, they have been under scrutiny over the years.
One line of scrutiny comes from the consistency with the 4d ${\cal N}=1$ SUGRA description. 
This issue, however, has been resolved in \cite{Bergshoeff:2015jxa} for the anti D3 uplift and in \cite{Parameswaran:2007kf, Achucarro:2006zf} for the BKQ scenario.
In the latter case refinements are needed either by retracting NP effects and adding instead $\alpha'$ and string loop corrections to the K\"ahler potential \cite{Parameswaran:2007kf}
or by adding light $U(1)$ charged matter to the magnetised D7-branes in such a way as to render the NP potential invariant under the shift symmetry \cite{Achucarro:2006zf}. Even though both scenarios appear to break SUSY explicitly, they can be realised as spontaneous SUSY breaking in the 4d ${\cal N}=1$ SUGRA description hence without the consistency issue.
Another line of scrutiny is specific to the anti D3 uplift and asks if the backreaction of anti D3-branes actually leads to the warped 4d dS space in type IIB flux backgrounds \cite{Bena:2009xk}. (See also \cite{McGuirk:2009xx, Gautason:2013zw}.) This issue has been studied locally in the Klebanov-Strassler (KS) throat, which is the prototypical flux background \cite{Klebanov:2000hb}, without the NP effects, as they are not essential to this issue. It was found that the backreaction of anti D3-branes develops singular 3-form fluxes and thus appears to infringe KKLT's uplifting scenario. 
Even if anti D3-branes puff up into NS5-branes via the mechanism suggested by Kachru, Pearson and Verlinde (KPV) \cite{Kachru:2002gs}, it was argued in the last paper of \cite{Bena:2009xk} that a perturbative instability is present as opposed to KPV's claim, posing the challenge to the anti D3 uplift.\footnote{A recent summary of this issue can be found in \cite{Polchinski:2015bea} where arguments for KKLT's uplifting scenarios are given and an optimism is expressed.}

Meanwhile, it is well-known that dS vacua, irrespective of the KKLT construction, are hard to realise in string theory first and foremost due to the constraint  spelled out by Gibbons, Maldacena and Nu\~nez (GMN) \cite{Gibbons:2003gb, Maldacena:2000mw}. The stringy effects, notably, orientifold planes and other (higher derivative) $\alpha'$ corrections as well as loop corrections, can in principle circumvent GMN's no-go argument against dS vacua. However, evidence so far has been  
challenging the existence of dS vacua in heterotic, type II and M theories \cite{Green:2011cn, Gautason:2012tb, Dasgupta:2014pma, Kutasov:2015eba, Quigley:2015jia}.

The common approach to the KKLT flux compactification is to utilise the 4d effective ${\cal N}=1$ supergravity (SUGRA) description. In simplest circumstances, as demonstrated by GKP \cite{Giddings:2001yu}, the 4d effective theory can be derived from type IIB SUGRA by dimensional reduction.
However, Einstein's equations of the 4d effective theory are only a part of type IIB SUGRA equations. 
Not only do they have to be supplemented by the tadpole cancellation condition, but it is also important to take into account the GMN constraint \cite{Gibbons:2003gb, Maldacena:2000mw}. In particular, we wish to shed light on the fact that it is not enough to evade their no-go theorem in order to realise dS vacua and, especially, the uplifting scenarios to dS vacua are in serious tension with the GMN constraint.
This is along the line of claims made in \cite{Green:2011cn, Gautason:2012tb, Dasgupta:2014pma, Kutasov:2015eba, Quigley:2015jia}. 
Our work can be thought of as a paraphrasing of \cite{Dasgupta:2014pma} and makes more manifest contact with the 4d effective SUGRA approach commonly exploited in the majority of the literature on KKLT and landscape. It can also be regarded as an emphasis on the use of consistent truncation in the spirit of \cite{Buchel:2003js, Quigley:2015jia}\footnote{We thank Daniel Elander and Evan McDonough for suggesting this viewpoint.}

The organisation of the paper is as follows. In section \ref{sec:equations} we review and elaborate on the link between the 4d effective theory and the type IIB SUGRA. In particular, we elucidate how the (trace of) Einstein's equations in the 4d effective theory can be extracted from type IIB SUGRA equations. In section \ref{sec:constraints} we discuss the two constraints advertised in the abstract. Based on these two constraints, we argue that the uplifting scenarios for dS vacua pose a serious obstacle to the evasion of GMN's no-go theorem.   
We briefly summarise the results and make short discussions in section \ref{sec:discussions}.

\section{The 4d effective SUGRA and type IIB equations}
\label{sec:equations}

The main purpose of this section is to link the 4d effective SUGRA approach with the 10d type IIB SUGRA in flux compactifications. Since it has been over a decade since KKLT,  their link has obviously been known. Nevertheless, we feel that the 4d effective theory has been trusted in isolation more than it is credited, and we wish to underscore the importance of  the constraints imposed by the type IIB equations to be taken into account.
We hope that our presentation here elucidates this point.

\subsection{The 4d effective SUGRA equations}
\label{sec:4dSUGRA}

The 4d effective ${\cal N}=1$ SUGRA theory is specified by the superpotential $W$ and the K\"ahler potential $K$ in terms of which the scalar potential takes the form:
\begin{align}
V={1\over 2\kappa_{10}^2}e^K\left(G^{a\bar{b}}D_aW\overline{D_bW}-3|W|^2\right)
\end{align}
where $\kappa_{10}^2$ (multiplied by the closed string coupling $g_s^2$) denotes 10d Newton's constant, $D_aW=\del_aW+W\del_aK$ and $G_{a\bar{b}}=\del_a\del_{\bar{b}}K$, and the indices $a$, $b$ label all superfields corresponding to K\"ahler, complex structure and axio-dilaton moduli of CY compactifications. Provided that the radius of the CY 3-fold is sufficiently large and neglecting (higher derivative) $\alpha'$ corrections,\footnote{Even for the large-radius CY cases, the $\alpha'$ corrections can be the dominant contribution \cite{Balasubramanian:2004uy, Westphal:2006tn, Parameswaran:2007kf}.} the K\"ahler potential $K$ following from the dimensional reduction is given by 
\cite{Giddings:2001yu}
\begin{align}
K=-3\ln\left[-i(\rho-\bar{\rho})\right]-\ln\left[-i(\tau-\bar{\tau})\right]-\ln\left(-i\int_{{\rm CY}_3}\Omega\wedge\bar{\Omega}\right)
\end{align}
where $\rho$ is the K\"ahler modulus whose imaginary part ${\rm Im}\rho\propto R_{\rm CY}^4$ with $R_{\rm CY}$ being the CY radius, $\tau=C_0+ie^{-\phi}$ is the axio-dilaton field with the RR 0-form field $C_0$ and the dilaton $\phi$, and $\Omega$ is the holomorphic $(3,0)$ form on the CY 3-fold which depends on the complex structure moduli. In this paper we restrict ourselves to the case with one K\"ahler modulus for brevity, but it is straightforward to generalise it to the multi-moduli case. Meanwhile, the superpotential $W$ is related to the RR and NSNS 3-form field strengths $F_3=dC_2$ and $H_3=dB_2$ by \cite{Gukov:1999ya}
\begin{align}
W=\int_{{\rm CY}_3} G_3\wedge \Omega
\end{align}
where $G_3=F_3-\tau H_3$. Moreover, the scalar potential can be identified with
\begin{align}
V={1\over\left({\rm Im}\rho\right)^3}\int_{{\rm CY}_3}d^6y\sqrt{\tilde{g}}\left[{|G_3|^2\over 24\kappa_{10}^2{\rm Im}\tau}
+T_3\rho^{\rm loc}_3\right]\equiv {1\over\left({\rm Im}\rho\right)^3}\int_{{\rm CY}_3}d^6y\sqrt{\tilde{g}}\,{\cal V}
\label{4dpotential}
\end{align}
where $\tilde{g}_{mn}$ and $y_m$'s are the metric and coordinates of the CY 3-fold, respectively. $T_3$ is the D3-brane tension, and $\rho^{\rm loc}_3$ is the D3 charge density from localised sources. This was derived in \cite{Giddings:2001yu} when the warp factor $A$, to be defined in \eqref{ansatz}, is constant and the RR 5-form field strength $\tilde{F}_5=F_5-\half C_2\wedge H_3+\half B_2\wedge F_3$ is absent in the type IIB background.

In the following we shall only be interested in the trace of Einstein's equations. To this end we note the trace of Einstein's equations of the 4d effective theory: 
\begin{align}
R_4=4G_NV
\label{4dtrace}
\end{align}
where $R_4$ is the Ricci scalar of the 4d spacetime and $G_N$ is 4d Newton's constant. The positive minima of the potential $V$ imply the existence of (meta-)stable dS vacua. As mentioned in the introduction, KKLT provided a way to construct such vacua: First, by adding a NP potential $W_{\rm np}\propto e^{ia\rho}$, with $a$ being constants, induced either by Euclidean D3-branes or gaugino condensation on D7-branes wrapping a 4-cycle in the CY 3-fold, the potential $V$ develops negative minima yielding SUSY AdS vacua. Then by further adding anti D3-branes $V_{\overline{D3}}\propto 2T_3/({\rm Im}\rho)^3$ and breaking SUSY, the minima of the potential can be raised to positive values uplifting the AdS vacua to dS vacua.

In the next sections we first derive the trace of 4d Einstein's equations \eqref{4dtrace} from the type IIB SUGRA equations
and then study it in conjunction with the GMN constraint.

\subsection{The type IIB equations}
\label{sec:IIBSUGRA}

As alluded to in the previous section, 
we are concerned with warped compactifications on Calabi-Yau (CY) 3-folds with the ansatz \cite{Giddings:2001yu}:
\be
ds_{10}^2=e^{2A(y)}g_{\mu\nu}dx^{\mu}dx^{\nu}+e^{-2A(y)}\tilde{g}_{mn}dy^mdy^n\equiv G_{MN}dX^MdX^N
\label{ansatz}
\ee
where Greek letters are used for the 4d spacetime $\mu, \nu=0,1,2,3$, whereas Latin letters for the 6d internal space $m, n=5, \cdots, 9$. The capital Latin letters $M, N$ collectively denote the indices for the 4d spacetime and the 6d internal space. The metric $g_{\mu\nu}$ is that of the 4d spacetime, whereas the metric $\tilde{g}_{mn}$ is that of the CY 3-fold. The warp factor $A(y)$ only depends on the coordinates $y^m$ of the internal space. 

The 10d Einstein's equations are given by
\begin{align}
R_{MN}=\kappa_{10}^2\left(T_{MN}-{1\over 8}g_{MN}T_L^L\right)
\end{align}
where $R_{MN}$ is the Ricci tensor constructed from the metric $G_{MN}$ and the energy-momentum (EM) tensor $T_{MN}$ has contributions from the 3-form $G_3=F_3-\tau H_3$, the 5-form $\tilde{F}_5=F_5-\half C_2\wedge H_3+\half B_2\wedge F_3$ and the axio-dilaton $\tau$. The 5-form is taken to be of the form
\be
\tilde{F}_5=(1+\ast)\left[\del_m\alpha dy^m\wedge dx^0\wedge dx^1\wedge dx^2\wedge dx^3\right]\ .
\ee
As given in \cite{Giddings:2001yu}, the trace of the 4d spacetime components yields 
\begin{align}
e^{2A}{\widetilde{\nabla}}^2A={1\over 4}e^{-2A}R_4+\left[{|G_3|^2\over 48{\rm Im}\tau}+{\kappa_{10}^2\over 8}\left(T^m_m-T^{\mu}_{\mu}\right)^{\rm loc}\right]+{e^{-8A}\over 4}\del_m\alpha\del^m\alpha
\label{4dEinstein}
\end{align}
where $R_4$ is the Ricci scalar for the unwarped 4d metric $g_{\mu\nu}$, $(T_{N}^M)^{\rm loc}$ is the energy-momentum tensor of the local sources and a tilde denotes use of the metric $\tilde{g}_{mn}$ (e.g., $\del^m = e^{2A}\del^{\tilde{m}}$). This equation, when integrated over the CY 3-fold, gives rise to the constraint derived by GMN.
In the meantime, the trace of the 6d internal space components yields
\begin{align}
e^{2A}{\widetilde{\nabla}}^2A={|G_3|^2\over 48{\rm Im}\tau}
+{\kappa_{10}^2\over 8}\left[{1\over 3}T^m_m-T^{\mu}_{\mu}\right]^{\rm loc}\!\!\!\!+{4\over 3}e^{2A}(\widetilde{\nabla}A)^2
+{e^{-8A}\over 6}\del_m\alpha\del^m\alpha
+{\del_m\tau\del^m\bar{\tau}\over 12({\rm Im}\tau)^2}\ .
\label{6dEinstein}
\end{align}
We now wish to relate these two equations to the trace of Einstein's equations \eqref{4dtrace} in the 4d effective theory. As it turns out, one needs to consider the linear combination $\eqref{4dEinstein}-3\times\eqref{6dEinstein}$ of these two equations. This yields
\begin{align}
{1\over 4}e^{-2A}R_4
=\kappa_{10}^2{\cal V}-\kappa_{10}^2\left[{1\over 4}(T^{\mu}_{\mu})^{\rm loc}+T_3\rho_3^{\rm loc}\right]+{e^{-8A}\over 4}\del_m\alpha\del^m\alpha+{\del_m\tau\del^m\bar{\tau}\over 4({\rm Im}\tau)^2}+e^{4A}\widetilde{\nabla}^2e^{-2A}
\label{4dequation}
\end{align}
where ${\cal V}$ is the scalar potential density defined in \eqref{4dpotential}. Since the EM tensor of the localised sources takes the form $T_{\mu\nu}^{\rm loc}=-e^{2A}g_{\mu\nu}\sum_p N_pT_p\delta(\Sigma_{p-3})$, where $\Sigma_{p-3}$ is the $(p-3)$-cycle of the CY 3-fold which a $p$-brane wraps and $N_p$ is the number of (anti-)D$p$-branes and O$p$-planes, its trace is given by $(T^{\mu}_{\mu})^{\rm loc}=-4\sum_p N_pT_p\delta(\Sigma_{p-3})$. Integrating this equation over the CY 3-fold when the warp factor $A$ is approximated by a constant,\footnote{It is not essential to consider the constant warp factor $A$, but we find it more clarifying to present the result for this simpler case.} one indeed finds \eqref{4dtrace} with a suitable generalisation which includes localised D-brane and O-plane sources, the 5-form flux and the axio-dilaton contributions:
\begin{align}
R_4=4\kappa_4^2V+4\kappa_4^2\left[\sum_p N_pT_pV_{p-3}-T_3Q_3^{\rm loc}\right]
+\int_{{\rm CY}_3}\!\!\!{d^6y\sqrt{\tilde{g}}\over V_{\rm CY}}\left[e^{-4A}(\widetilde{\nabla}\alpha)^2+{e^{4A}|\widetilde{\nabla}\tau|^2\over ({\rm Im}\tau)^2}\right]
\label{4dfrom10d}
\end{align}
where $\kappa_4^2=e^{2A}\kappa_{10}^2/V_{\rm CY}$ with the CY volume $V_{\rm CY}$, the K\"ahler modulus $\rho$ is set to be a constant, $V_{p-3}$ is the volume of the $(p-3)$-cycle and we used $\del^m=e^{2A}\del^{\tilde{m}}$.
Note that as one can see from the localised source term, $N_{\bar{3}}$ anti D3-branes increase the cosmological constant by $2\kappa_4^2N_{\bar{3}}T_3$ (justifiably neglecting the change of the 5-form flux and the potential $V$). Reinstating the K\"ahler modulus dependence, this is of course the uplifting potential $V_{\overline{D3}}\propto 2T_3/({\rm Im}\rho)^3$ considered in KKLT.

\section{The constraints on the uplifts to dS vacua}
\label{sec:constraints}

To elucidate how the GMN constraint \cite{Gibbons:2003gb, Maldacena:2000mw} might obstruct the uplifting scenarios, we shall have a closer look at the two equations \eqref{4dEinstein} and \eqref{4dequation} in the type IIB SUGRA. Upon integration over the CY 3-fold, the former leads to the GMN constraint, whereas the latter, as elaborated above, yields the (trace of) Einstein's equations \eqref{4dfrom10d} in the 4d effective ${\cal N}=1$ SUGRA. As given in \cite{Giddings:2001yu} from \eqref{4dEinstein}, the GMN constraint reads\footnote{Upon integration the warp factor $A$ is assumed to be smooth on the CY 3-fold. However, to be rigorous, one needs to take proper care of brane source singularities which may require the M-theory lift \cite{Dasgupta:2014pma}.}
\begin{align}
\int_{{\rm CY}_3}d^6y\sqrt{\tilde{g}}e^{-4A}\biggl[R_4+2\kappa_{10}^2e^{2A}{\cal V}+2\kappa_{10}^2e^{2A}T_4^{\rm loc}+e^{-4A}(\widetilde{\nabla}\alpha)^2\biggr]=0\label{constraint}
\end{align}
where $T_4^{\rm loc}$ is the (partial) trace of the 10d EM tensor for the localised sources over the 4d spacetime and, using $(T^m_m)^{\rm loc}=\sum_p(3-p)N_pT_p\delta(\Sigma_{p-3})$, it is given by
\begin{align}
T_4^{\rm loc}=\sum_p{7-p\over 4}N_pT_p\delta(\Sigma_{p-3})-T_3\rho^{\rm loc}_3\ .
\label{T4}
\end{align}
In order to discuss how constraining the GMN constraint \eqref{constraint} may be for the uplifting scenarios, instead of considering the 4d equation in the form \eqref{4dfrom10d}, it is more convenient to integrate \eqref{4dequation} to 
\begin{align}
\int_{{\rm CY}_3}d^6y\sqrt{\tilde{g}}e^{-6A}\left[-R_4+4\kappa_{10}^2e^{2A}{\cal V}+4\kappa_{10}^2e^{2A}T^{\rm loc}+e^{-4A}(\widetilde{\nabla}\alpha)^2+{e^{4A}|\widetilde{\nabla}\tau|^2\over ({\rm Im}\tau)^2}\right]=0
\label{constraint4d}
\end{align}
where we introduced the notation $T^{\rm loc}$ for the localised source contributions given by
\begin{align}
T^{\rm loc}=\sum_p N_pT_p\delta(\Sigma_{p-3})-T_3\rho_3^{\rm loc}\ .
\label{T}
\end{align}
On top of these constraints, one needs to take into account the tadpole cancellation condition $\int_{{\rm CY}_3}H_3\wedge F_3+2\kappa_{10}^2T_3Q_3^{\rm loc}=0$, which follows from the 5-form Bianchi identity, and the 3-form and the axio-dilaton equations. However, these latter equations are not of our focus in this paper.

We now focus on the two constraints, the GMN constraint \eqref{constraint} and the 4d theory constraint \eqref{constraint4d}, and discuss them in view of the uplifting scenarios of KKLT and BKQ. It is assumed that the dilaton is fixed at a weak coupling value and the K\"ahler modulus at a large radius so that the perturbative description and the $\alpha'$ derivative expansion make sense. First, we can incorporate the NP effects, $\alpha'$ and loop corrections by shifting the potential ${\cal V}$:
\be
{\cal V}\quad\longrightarrow\quad 
{\cal V}+{\cal V}_{\rm np}'+{\cal V}_{\alpha'+g_s}'
\,\,\,{\rm in}\,\,\,\eqref{constraint}\ , \qquad
{\cal V}+{\cal V}_{\rm np}+{\cal V}_{\alpha'+g_s}
\,\,\,{\rm in}\,\,\,\eqref{constraint4d}\ .
\ee
Note that depending on which sources these corrections originate from, they generally differ in the two constraints.
Meanwhile, the D-term potential in BKQ coming from the magnetic flux $B$ on a wrapped D7-brane shifts $(T^{D7})^{\mu}_{\mu}$ by $-4T_7B^2\delta(\Sigma_4)/({\rm Im}\rho)^3$, whereas $(T^{D7})^{m}_{m}$ is unchanged. This correction thus shifts both $T^{\rm loc}$ and $T_4^{\rm loc}$ by the same amount $4T_7B^2\delta(\Sigma_4)/({\rm Im}\rho)^3$. In the case that D7-branes and O7-planes induce D3 charges via the CS couplings, their effects are incorporated through the D3 charge density $\rho_3^{\rm loc}$ (and its backreaction to $G_3$ and $\alpha$) and therefore shifts ${\cal V}$, $T_4^{\rm loc}$ and $T^{\rm loc}$ accordingly. There are also related shifts to $T_4^{\rm loc}$ and $T^{\rm loc}$ due to higher curvature $\alpha'$ corrections to the DBI action of D7-branes and O7-planes.

In the first step of KKLT, since the vacua are AdS, the 4d Ricci scalar $R_4$ is negative. Thus the two constraints require that
\begin{align}
&\eqref{constraint}\quad\stackrel{AdS}{\Longrightarrow}\quad
2\kappa_{10}^2e^{2A}\left({\cal V}+{\cal V}_{\rm np}'\right)+2\kappa_{10}^2e^{2A}T_4^{\rm loc}+e^{-4A}(\widetilde{\nabla}\alpha)^2>0\ ,\label{GMNAdS}\\
&\eqref{constraint4d}\quad\stackrel{AdS}{\Longrightarrow}\quad
4\kappa_{10}^2e^{2A}\left({\cal V}+{\cal V}_{\rm np}\right)+4\kappa_{10}^2e^{2A}T^{\rm loc}+e^{-4A}(\widetilde{\nabla}\alpha)^2+{e^{4A}|\widetilde{\nabla}\tau|^2\over ({\rm Im}\tau)^2}<0\ .
\label{4dEFTAdS}
\end{align}
In particular, the NP potential ${\cal V}_{\rm np}$ is the contribution which drives down the minima of the potential to negative values. Irrespective of the details of how these two conditions are realised, assuming that they are satisfied, we now ask if adding anti D3-branes can uplift the AdS vacua in consistent with the two constraints. We first note that (anti-)D3-branes as well as O3-planes make same contributions to $T_4^{\rm loc}$ and $T^{\rm loc}$, as one can see from \eqref{T4} and \eqref{T}. This implies that the anti D3-branes increase both of them by the same amount 
\begin{align}
\Delta T_4^{\rm loc}=\Delta T^{\rm loc}= 2N_{\bar{3}}T_3\delta^6(y-y_i)>0 \label{antiD3}
\end{align}
where $y_i$ collectively denotes the locations of anti D3-branes.  
In accordance with KKLT's 4d effective theory analysis, this contribution can uplift the potential to achieve
\begin{align}
\hspace{-.4cm}
\eqref{constraint4d}\,\,\,\stackrel{dS}{\Longrightarrow}\,\,\,
4\kappa_{10}^2e^{2A}\left({\cal V}+{\cal V}_{\rm np}\right)+4\kappa_{10}^2e^{2A}\left(T^{\rm loc}+\Delta T^{\rm loc}\right)+e^{-4A}(\widetilde{\nabla}\alpha)^2+{e^{4A}|\widetilde{\nabla}\tau|^2\over ({\rm Im}\tau)^2}>0\ .
\label{antiD34dconstraint}
\end{align}
However, provided that the anti D3-branes energy is the dominant source of the change, it seems impossible to satisfy the GMN constraint \eqref{constraint}, since \eqref{GMNAdS} and \eqref{antiD3} together yield
\begin{align}
\kappa_{10}^2e^{2A}\left({\cal V}+{\cal V}_{\rm np}'\right)+2\kappa_{10}^2e^{2A}\left(T_4^{\rm loc}+\Delta T_4^{\rm loc}\right)+e^{-4A}(\widetilde{\nabla}\alpha)^2>0
\end{align}
which implies a negative cosmological constant, $R_4<0$, in contradiction with the positive cosmological constant \eqref{antiD34dconstraint}.

In the case of the BKQ D-term uplift, we consider the refined version \cite{Parameswaran:2007kf} in which the NP potential is absent and instead $\alpha'$ and loop corrections are present.\footnote{We will not discuss the other type of the refined BKQ scenario \cite{Achucarro:2006zf} which involves light matter on D7-branes and requires sophisticated arrangements of matter. } As in the KKLT construction, it starts with AdS vacua and thus the two constraints again require that\footnote{The Ricci scalar $R_4$ is in the 10d Einstein frame, and here it is implicitly Weyl-scaled to the 4d Einstein frame. The $\alpha'+g_s$ corrections have contributions from the $R_4$ Weyl scaling and the 3-form fluxes in ${\cal V}$ \cite{Becker:2002nn}. These corrections thus appear differently in the two constraints.}
\begin{align}
&\eqref{constraint}\quad\stackrel{AdS}{\Longrightarrow}\quad
2\kappa_{10}^2e^{2A}\left({\cal V}+{\cal V}_{\alpha'+g_s}'\right)+2\kappa_{10}^2e^{2A}T_4^{\rm loc}+e^{-4A}(\widetilde{\nabla}\alpha)^2>0\ ,\label{GMNAdS2}\\
&\eqref{constraint4d}\quad\stackrel{AdS}{\Longrightarrow}\quad
4\kappa_{10}^2e^{2A}\left({\cal V}+{\cal V}_{\alpha'+g_s}\right)+4\kappa_{10}^2e^{2A}T^{\rm loc}+e^{-4A}(\widetilde{\nabla}\alpha)^2+{e^{4A}|\widetilde{\nabla}\tau|^2\over ({\rm Im}\tau)^2}<0\ .
\end{align}
Similar to the previous case, the $\alpha'$ and loop corrections ${\cal V}_{\alpha'+g_s}$ drive down the minima of the potential to negative values. Again, we assume that these constraints are obeyed. We now add a magnetised D7-brane which induces the D-term. (The magnetised D7-brane has to be accompanied by an O7-plane and three D7-branes well separated from it.) In this case, as explained above, both $T^{\rm loc}$ and $T_4^{\rm loc}$ are shifted by $\Delta T^{\rm loc}=\Delta T_4^{\rm loc} \sim 4T_7B^2\delta(\Sigma_4)$ (suppressing the K\"ahler modulus $\rho$ dependence), while the potential ${\cal V}$ remains the same.
Thus, in accordance with BKQ's 4d effective theory analysis, the magnetised D7-brane can uplift the potential to achieve
\begin{align}
\hspace{-.45cm}
\eqref{constraint4d}\,\stackrel{dS}{\Longrightarrow}\,
4\kappa_{10}^2e^{2A}\left({\cal V}+{\cal V}_{\alpha'+g_s}\right)+4\kappa_{10}^2e^{2A}\left(T^{\rm loc}+\Delta T^{\rm loc}\right)+e^{-4A}(\widetilde{\nabla}\alpha)^2+{e^{4A}|\widetilde{\nabla}\tau|^2\over ({\rm Im}\tau)^2}>0\ .
\label{mD74dconstraint}
\end{align}
However, provided that the magnetised D7-brane D-term energy is the dominant source of the change, it seems again impossible to satisfy the GMN constraint:
\begin{align}
\kappa_{10}^2e^{2A}\left({\cal V}+{\cal V}_{\alpha'+g_s}'\right)+2\kappa_{10}^2e^{2A}\left(T_4^{\rm loc}+\Delta T_4^{\rm loc}\right)+e^{-4A}(\widetilde{\nabla}\alpha)^2>0\ ,
\end{align} 
which is again in contradiction with the positive cosmological constant \eqref{mD74dconstraint}.

In hindsight these results may have been anticipated more evidently, as a different linear combination of the two trace equations $\eqref{4dEinstein}-\eqref{6dEinstein}$ reads
\begin{align}
\hspace{-.1cm}
R_4=\kappa_{10}^2e^{2A}\sum_p{p-3\over 3}T_p\delta(\Sigma_{p-3})
+{1\over 3}e^{-4A}\!\left( \del_m e^{4A}+\del_m\alpha\right)\!\left( \del^{\tilde{m}}e^{4A}-\del^{\tilde{m}}\alpha\right)
+{e^{4A}|\widetilde{\nabla}\tau|^2\over 3({\rm Im}\tau)^2}
\label{difference}
\end{align}
where the localised source contributions only come from $(T_m^m)^{\rm loc}$ to which gauge fluxes on D7-branes do not contribute.
Note that there are NP and $\alpha'+g_s$ corrections present on the RHS, but we omitted them just to focus on the following point. 
Neither do anti D3-branes nor the magnetised D7-brane D-term contribute {\it directly} to the 4d cosmological constant, hence the uplifts may be inhibited. 

\subsection{The case of non-Ricci flat 6d manifolds}
\label{sec:nonCY}

It could happen that the 6d compact manifolds cannot be Ricci flat, $\tilde{R}_6\ne 0$, for one reason or another, in particular, in the presence of D7-branes and O7-planes and thus fail to be CY \cite{Dasgupta:2008hw}. This affects the internal components of the Einstein's equations and adds the 6d curvature contribution to \eqref{6dEinstein}: 
\begin{align}
e^{2A}{\widetilde{\nabla}}^2A=-{1\over 6}e^{2A}\tilde{R}_6+{|G_3|^2\over 48{\rm Im}\tau}
+{\kappa_{10}^2\over 8}\left[{1\over 3}T^m_m-T^{\mu}_{\mu}\right]^{\rm loc}\!\!\!\!+{4\over 3}e^{2A}(\widetilde{\nabla}A)^2
+{e^{-8A}\over 6}\del_m\alpha\del^m\alpha\ .
\label{nonRicci6dEinstein}
\end{align}
This change is reflected in \eqref{difference} as
\begin{align}
R_4={\rm RHS\,\, of\,\,}\eqref{difference}-{2\over 3}e^{4A}\tilde{R}_6\ .
\end{align}
The presence of $\tilde{R}_6$, however, has little significance to the above arguments for the challenge the uplifts face: We start with SUSY AdS vacua for which $R_4 < 0$. Regardless of the presence of $\tilde{R}_6$, what matters in the uplifting scenarios is whether anti D3-branes or the magnetised D7-brane can increase $R_4$ to a positive value or not. As we have just seen, the answer to this question is negative.  

These considerations suggest that the uplifting scenarios obstruct the evasion of GMN's no-go theorem more strongly than one might have assumed.

\section{Discussions}
\label{sec:discussions}

We revisited the issue of uplifting the potential to dS vacua in type IIB flux compactifications of KKLT. We shed light on some tension between two constraints on dS vacua in type IIB string theory. 
One is the well-known GMN's no-go theorem that can by itself be evaded. The other follows from the (generalisation of) Einstein's equations in the 4d effective theory which has, however, been hardly discussed in connection with the GMN constraint. We argued that the Einstein's equations of the 4d effective theory are only a part of type IIB equations and emphasised the importance of the GMN constraint to be taken into account on top of the tadpole cancellation condition. 
In particular, in addition to the challenges previously posed, it is suggested that the uplifting scenarios obstruct the evasion of the no-go theorem more invincibly than one might have assumed.

In order for dS vacua to exist, from \eqref{constraint} and \eqref{constraint4d} with $\tilde{R}_6$ included,  the following two constraints need to be satisfied:
\begin{align}
&
2\kappa_{10}^2e^{2A}\left({\cal V}+{\cal V}_{\rm np}'+{\cal V}_{\alpha'+g_s}'\right)+2\kappa_{10}^2e^{2A}T_4^{\rm loc}\!+e^{-4A}(\widetilde{\nabla}\alpha)^2<0\ ,\label{GMNdS}\\
&
4\kappa_{10}^2e^{2A}\left({\cal V}+{\cal V}_{\rm np}+{\cal V}_{\alpha'+g_s}\right)+4\kappa_{10}^2e^{2A}T^{\rm loc}\!+e^{-4A}(\widetilde{\nabla}\alpha)^2+{e^{4A}|\widetilde{\nabla}\tau|^2\over ({\rm Im}\tau)^2}
-2e^{4A}\tilde{R}_6>0\ .
\label{4dtheoryconstraint}
\end{align}
More properly, these constraints should be considered in the integrated form.

Our statements are restricted to the brane-based uplifting scenarios in which the second constraint can be satisfied, but it is very hard, if not impossible, to simultaneously obey the first GMN constraint.
It is weaker than to claim that these two constraints can never be compatible with each other, but without uplifting mechanisms it is much harder to look for meta-stable dS vacua consistent with both constraints.\footnote{There have been important developments in the explicit construction of meta-stable dS vacua in type IIA string theory \cite{Hertzberg:2007wc, Silverstein:2007ac, Andriot:2010ju, Haque:2008jz} which in particular involve the negative curvature $\tilde{R}_6 < 0$ contribution \cite{Douglas:2010rt}.}

In this connection, we would like to end our discussions with a comment on the search for dS vacua at $V > 0$ without uplifts in the no-scale model such as the work \cite{Saltman:2004sn}.
For this class of dS vacua (whose K\"ahler moduli yet to be fixed), we first note that the GMN constraint \eqref{GMNdS} seems to be violated, even though the 4d theory constraint \eqref{4dtheoryconstraint} is fulfilled. With the understanding that the constraints are integrated over 6d manifolds, this statement withstands even in the presence of D7-branes and O7-planes when the 7-branes tadpole cancellation is properly taken into account. 
However, the emphasis should be given to the fact that these vacua are meta-stable local minima with the complex structure and the dilaton moduli fixed. This might suggest a construction of dS vacua reversing the two steps in the KKLT construction: Starting with these tentative dS vacua, the addition of NP potentials and/or $\alpha'+g_s$ corrections could in principle render the GMN constraint \eqref{GMNdS} consistent, while the K\"ahler moduli settle to fixed values and the complex structure and the dilaton moduli undergo only small shifts. Although the cosmological constants would be reduced from those of the tentative dS vacua, they could remain positive, providing in principle  putative meta-stable dS vacua.

\section*{Acknowledgments}

We would especially like to thank Keshav Dasgupta and Evan McDonough for illuminating discussions and the comments on the manuscript, Daniel Elander for pointing out related relevant works and Oscar Loaiza-Brito for helpful discussions. NCB would like to thank the support of DAIP and PRODEP.
This work was supported in part by the National Research Foundation of South Africa and DST-NRF Centre of Excellence in Mathematical and Statistical Sciences (CoE-MaSS).
Opinions expressed and conclusions arrived at are those of the authors and are not necessarily to be attributed to the NRF or the CoE-MaSS.

\if{
\appendix

\section{Formal expansion}
\label{app:expand_S}
\fi


\begin{thebibliography}{99}

\bibitem{Linde:2015edk} 
  A.~Linde,
  ``A brief history of the multiverse,''
  arXiv:1512.01203 [hep-th].

\bibitem{Susskind:2003kw} 
  L.~Susskind,
  ``The Anthropic landscape of string theory,''
  In *Carr, Bernard (ed.): Universe or multiverse?* 247-266
  [hep-th/0302219].

\bibitem{Bousso:2000xa} 
  R.~Bousso and J.~Polchinski,
  ``Quantization of four form fluxes and dynamical neutralization of the cosmological constant,''
  JHEP {\bf 0006}, 006 (2000)
  doi:10.1088/1126-6708/2000/06/006
  [hep-th/0004134].

\bibitem{Kachru:2003aw} 
  S.~Kachru, R.~Kallosh, A.~D.~Linde and S.~P.~Trivedi,
  ``De Sitter vacua in string theory,''
  Phys.\ Rev.\ D {\bf 68}, 046005 (2003)
  doi:10.1103/PhysRevD.68.046005
  [hep-th/0301240].
  
\bibitem{Giddings:2001yu} 
  S.~B.~Giddings, S.~Kachru and J.~Polchinski,
  ``Hierarchies from fluxes in string compactifications,''
  Phys.\ Rev.\ D {\bf 66}, 106006 (2002)
  doi:10.1103/PhysRevD.66.106006
  [hep-th/0105097].


\bibitem{DeWolfe:2005uu} 
  O.~DeWolfe, A.~Giryavets, S.~Kachru and W.~Taylor,
  ``Type IIA moduli stabilization,''
  JHEP {\bf 0507}, 066 (2005)
  doi:10.1088/1126-6708/2005/07/066
  [hep-th/0505160].
  
  
\bibitem{Grana:2005jc} 
  M.~Gra\~na,
  ``Flux compactifications in string theory: A Comprehensive review,''
  Phys.\ Rept.\  {\bf 423}, 91 (2006)
  doi:10.1016/j.physrep.2005.10.008
  [hep-th/0509003].


\bibitem{Douglas:2003um} 
  M.~R.~Douglas,
  ``The Statistics of string / M theory vacua,''
  JHEP {\bf 0305}, 046 (2003)
  doi:10.1088/1126-6708/2003/05/046
  [hep-th/0303194].
  
  
\bibitem{Burgess:2003ic} 
  C.~P.~Burgess, R.~Kallosh and F.~Quevedo,
  ``De Sitter string vacua from supersymmetric D terms,''
  JHEP {\bf 0310}, 056 (2003)
  doi:10.1088/1126-6708/2003/10/056
  [hep-th/0309187];
  

\bibitem{Balasubramanian:2004uy} 
  V.~Balasubramanian and P.~Berglund,
  ``Stringy corrections to Kahler potentials, SUSY breaking, and the cosmological constant problem,''
  JHEP {\bf 0411}, 085 (2004)
  doi:10.1088/1126-6708/2004/11/085
  [hep-th/0408054];

\bibitem{Westphal:2006tn} 
  A.~Westphal,
  ``de Sitter string vacua from Kahler uplifting,''
  JHEP {\bf 0703}, 102 (2007)
  doi:10.1088/1126-6708/2007/03/102
  [hep-th/0611332].
  
   
\bibitem{Brustein:2004xn} 
  R.~Brustein and S.~P.~de Alwis,
  ``Moduli potentials in string compactifications with fluxes: Mapping the discretuum,''
  Phys.\ Rev.\ D {\bf 69}, 126006 (2004)
  doi:10.1103/PhysRevD.69.126006
  [hep-th/0402088].

\bibitem{Balasubramanian:2005zx} 
  V.~Balasubramanian, P.~Berglund, J.~P.~Conlon and F.~Quevedo,
  ``Systematics of moduli stabilisation in Calabi-Yau flux compactifications,''
  JHEP {\bf 0503}, 007 (2005)
  doi:10.1088/1126-6708/2005/03/007
  [hep-th/0502058].
  

\bibitem{Bergshoeff:2015jxa} 
  E.~A.~Bergshoeff, K.~Dasgupta, R.~Kallosh, A.~Van Proeyen and T.~Wrase,
  ``$ \overline{\mathrm{D}3} $ and dS,''
  JHEP {\bf 1505}, 058 (2015)
  doi:10.1007/JHEP05(2015)058
  [arXiv:1502.07627 [hep-th]].
  
\bibitem{Parameswaran:2007kf} 
  S.~L.~Parameswaran and A.~Westphal,
  ``de Sitter string vacua from perturbative Kahler corrections and consistent D-terms,''
  JHEP {\bf 0610}, 079 (2006)
  doi:10.1088/1126-6708/2006/10/079
  [hep-th/0602253];
  ``Consistent de Sitter string vacua from Kahler stabilization and D-term uplifting,''
  Fortsch.\ Phys.\  {\bf 55}, 804 (2007)
  doi:10.1002/prop.200610374
  [hep-th/0701215].
  
\bibitem{Achucarro:2006zf} 
  A.~Achucarro, B.~de Carlos, J.~A.~Casas and L.~Doplicher,
  ``De Sitter vacua from uplifting D-terms in effective supergravities from realistic strings,''
  JHEP {\bf 0606}, 014 (2006)
  doi:10.1088/1126-6708/2006/06/014
  [hep-th/0601190].

   
  
\bibitem{Bena:2009xk} 
  I.~Bena, M.~Gra\~na and N.~Halmagyi,
  ``On the Existence of Meta-stable Vacua in Klebanov-Strassler,''
  JHEP {\bf 1009}, 087 (2010)
  doi:10.1007/JHEP09(2010)087
  [arXiv:0912.3519 [hep-th]];
  I.~Bena, M.~Gra\~na, S.~Kuperstein and S.~Massai,
  ``Polchinski-Strassler does not uplift Klebanov-Strassler,''
  JHEP {\bf 1309}, 142 (2013)
  doi:10.1007/JHEP09(2013)142
  [arXiv:1212.4828 [hep-th]];
  I.~Bena, M.~Gra\~na, S.~Kuperstein and S.~Massai,
  ``Giant Tachyons in the Landscape,''
  JHEP {\bf 1502}, 146 (2015)
  doi:10.1007/JHEP02(2015)146
  [arXiv:1410.7776 [hep-th]].
  
\bibitem{McGuirk:2009xx} 
  P.~McGuirk, G.~Shiu and Y.~Sumitomo,
  ``Non-supersymmetric infrared perturbations to the warped deformed conifold,''
  Nucl.\ Phys.\ B {\bf 842}, 383 (2011)
  doi:10.1016/j.nuclphysb.2010.09.008
  [arXiv:0910.4581 [hep-th]].
  
\bibitem{Gautason:2013zw} 
  F.~F.~Gautason, D.~Junghans and M.~Zagermann,
  ``Cosmological Constant, Near Brane Behavior and Singularities,''
  JHEP {\bf 1309}, 123 (2013)
  doi:10.1007/JHEP09(2013)123
  [arXiv:1301.5647 [hep-th]].
  
\bibitem{Klebanov:2000hb} 
  I.~R.~Klebanov and M.~J.~Strassler,
  ``Supergravity and a confining gauge theory: Duality cascades and chi SB resolution of naked singularities,''
  JHEP {\bf 0008}, 052 (2000)
  doi:10.1088/1126-6708/2000/08/052
  [hep-th/0007191].
  
\bibitem{Kachru:2002gs} 
  S.~Kachru, J.~Pearson and H.~L.~Verlinde,
  ``Brane / flux annihilation and the string dual of a nonsupersymmetric field theory,''
  JHEP {\bf 0206}, 021 (2002)
  doi:10.1088/1126-6708/2002/06/021
  [hep-th/0112197].
  
  
\bibitem{Polchinski:2015bea} 
  J.~Polchinski,
  ``Brane/antibrane dynamics and KKLT stability,''
  arXiv:1509.05710 [hep-th].
 
    
  
\bibitem{Gibbons:2003gb} 
  G.~W.~Gibbons,
  ``Thoughts on tachyon cosmology,''
  Class.\ Quant.\ Grav.\  {\bf 20}, S321 (2003)
  doi:10.1088/0264-9381/20/12/301
  [hep-th/0301117].
  
\bibitem{Maldacena:2000mw} 
  J.~M.~Maldacena and C.~Nunez,
  ``Supergravity description of field theories on curved manifolds and a no go theorem,''
  Int.\ J.\ Mod.\ Phys.\ A {\bf 16}, 822 (2001)
  doi:10.1142/S0217751X01003935, 10.1142/S0217751X01003937
  [hep-th/0007018].
  

\bibitem{Green:2011cn} 
  S.~R.~Green, E.~J.~Martinec, C.~Quigley and S.~Sethi,
  ``Constraints on String Cosmology,''
  Class.\ Quant.\ Grav.\  {\bf 29}, 075006 (2012)
  doi:10.1088/0264-9381/29/7/075006
  [arXiv:1110.0545 [hep-th]].
  
\bibitem{Gautason:2012tb} 
  F.~F.~Gautason, D.~Junghans and M.~Zagermann,
  ``On Cosmological Constants from alpha'-Corrections,''
  JHEP {\bf 1206}, 029 (2012)
  doi:10.1007/JHEP06(2012)029
  [arXiv:1204.0807 [hep-th]].
  
\bibitem{Dasgupta:2014pma} 
  K.~Dasgupta, R.~Gwyn, E.~McDonough, M.~Mia and R.~Tatar,
  ``de Sitter Vacua in Type IIB String Theory: Classical Solutions and Quantum Corrections,''
  JHEP {\bf 1407}, 054 (2014)
  doi:10.1007/JHEP07(2014)054
  [arXiv:1402.5112 [hep-th]].
    
\bibitem{Kutasov:2015eba} 
  D.~Kutasov, T.~Maxfield, I.~Melnikov and S.~Sethi,
  ``Constraining de Sitter Space in String Theory,''
  Phys.\ Rev.\ Lett.\  {\bf 115}, no. 7, 071305 (2015)
  doi:10.1103/PhysRevLett.115.071305
  [arXiv:1504.00056 [hep-th]].


\bibitem{Quigley:2015jia} 
  C.~Quigley,
  ``Gaugino Condensation and the Cosmological Constant,''
  JHEP {\bf 1506}, 104 (2015)
  doi:10.1007/JHEP06(2015)104
  [arXiv:1504.00652 [hep-th]].

\bibitem{Buchel:2003js} 
  A.~Buchel,
  ``On effective action of string theory flux compactifications,''
  Phys.\ Rev.\ D {\bf 69}, 106004 (2004)
  doi:10.1103/PhysRevD.69.106004
  [hep-th/0312076].
 

  
\bibitem{Gukov:1999ya} 
  S.~Gukov, C.~Vafa and E.~Witten,
  ``CFT's from Calabi-Yau four folds,''
  Nucl.\ Phys.\ B {\bf 584}, 69 (2000)
  Erratum: [Nucl.\ Phys.\ B {\bf 608}, 477 (2001)]
  doi:10.1016/S0550-3213(00)00373-4
  [hep-th/9906070].


\bibitem{Becker:2002nn} 
  K.~Becker, M.~Becker, M.~Haack and J.~Louis,
  ``Supersymmetry breaking and alpha-prime corrections to flux induced potentials,''
  JHEP {\bf 0206}, 060 (2002)
  doi:10.1088/1126-6708/2002/06/060
  [hep-th/0204254].


\bibitem{Dasgupta:2008hw} 
  K.~Dasgupta, P.~Franche, A.~Knauf and J.~Sully,
  ``D-terms on the resolved conifold,''
  JHEP {\bf 0904}, 027 (2009)
  doi:10.1088/1126-6708/2009/04/027
  [arXiv:0802.0202 [hep-th]].


\bibitem{Hertzberg:2007wc} 
  M.~P.~Hertzberg, S.~Kachru, W.~Taylor and M.~Tegmark,
  ``Inflationary Constraints on Type IIA String Theory,''
  JHEP {\bf 0712}, 095 (2007)
  doi:10.1088/1126-6708/2007/12/095
  [arXiv:0711.2512 [hep-th]].

\bibitem{Silverstein:2007ac} 
  E.~Silverstein,
  ``Simple de Sitter Solutions,''
  Phys.\ Rev.\ D {\bf 77}, 106006 (2008)
  doi:10.1103/PhysRevD.77.106006
  [arXiv:0712.1196 [hep-th]].

\bibitem{Andriot:2010ju} 
  D.~Andriot, E.~Goi, R.~Minasian and M.~Petrini,
  ``Supersymmetry breaking branes on solvmanifolds and de Sitter vacua in string theory,''
  JHEP {\bf 1105}, 028 (2011)
  doi:10.1007/JHEP05(2011)028
  [arXiv:1003.3774 [hep-th]].

\bibitem{Haque:2008jz} 
  S.~S.~Haque, G.~Shiu, B.~Underwood and T.~Van Riet,
  ``Minimal simple de Sitter solutions,''
  Phys.\ Rev.\ D {\bf 79}, 086005 (2009)
  doi:10.1103/PhysRevD.79.086005
  [arXiv:0810.5328 [hep-th]];
  U.~H.~Danielsson, S.~S.~Haque, G.~Shiu and T.~Van Riet,
  ``Towards Classical de Sitter Solutions in String Theory,''
  JHEP {\bf 0909}, 114 (2009)
  doi:10.1088/1126-6708/2009/09/114
  [arXiv:0907.2041 [hep-th]].
  U.~H.~Danielsson, P.~Koerber and T.~Van Riet,
  ``Universal de Sitter solutions at tree-level,''
  JHEP {\bf 1005}, 090 (2010)
  doi:10.1007/JHEP05(2010)090
  [arXiv:1003.3590 [hep-th]].

\bibitem{Douglas:2010rt} 
  M.~R.~Douglas and R.~Kallosh,
  ``Compactification on negatively curved manifolds,''
  JHEP {\bf 1006}, 004 (2010)
  doi:10.1007/JHEP06(2010)004
  [arXiv:1001.4008 [hep-th]].


\bibitem{Saltman:2004sn} 
  A.~Saltman and E.~Silverstein,
  ``The Scaling of the no scale potential and de Sitter model building,''
  JHEP {\bf 0411}, 066 (2004)
  doi:10.1088/1126-6708/2004/11/066
  [hep-th/0402135].

\end{thebibliography}
\end{document}